%% file: main.tex
\documentclass[sigconf]{aamas}

\usepackage{balance} 
\usepackage{tabularx}
\usepackage{tikz}
\usepackage{svg}
\usepackage{pgfplots}
\pgfplotsset{compat=1.18}
\usepackage{pgfplotstable}
\usetikzlibrary{matrix,positioning}
\usepgfplotslibrary{colorbrewer}
\pgfplotsset{compat = 1.15, cycle list/Set1-8} 
\usetikzlibrary{pgfplots.statistics, pgfplots.colorbrewer} 
\usepackage{pgfplotstable}
\usepackage{filecontents}
\usepackage{varwidth}
\usepackage{tabularx}
\usepackage{soul}
\usepackage[export]{adjustbox}
\usepackage{dblfloatfix}

\usetikzlibrary{pgfplots.groupplots}
\usetikzlibrary{shapes.geometric}
\usetikzlibrary{positioning,fit,shapes.geometric,backgrounds, calc}
\usetikzlibrary{arrows,decorations.markings}
\usetikzlibrary{shapes.arrows}
\usetikzlibrary{patterns}
\tikzset{textnode/.style={inner sep=0pt,outer sep=0,execute at begin node={\strut}}}
\tikzstyle{state} = [textnode,circle, draw, inner sep=0pt, outer sep=0]
\usepgfplotslibrary{groupplots}

\definecolor{ppt-purple}{RGB}{112, 48, 160}
\definecolor{ppt-blue}{RGB}{91, 155, 213}
\definecolor{ppt-orange}{RGB}{237, 125, 49}
\definecolor{ppt-gray}{RGB}{165, 165, 165}
\definecolor{ppt-yellow}{RGB}{255, 192, 0}
\definecolor{ppt-darkblue}{RGB}{68, 114, 196}
\definecolor{ppt-green}{RGB}{112, 173, 71}
\definecolor{ppt-red}{RGB}{255, 0, 0}

\pgfplotsset{every axis/.append style={
    xlabel={$x$},          
    ylabel={$y$},          
    label style={font=\sffamily},
    tick label style={font=\sffamily\scriptsize},
    xticklabel style = {font=\sffamily\scriptsize},
    title style = {font=\normalsize\sffamily},
    ylabel near ticks,
    y label style={font=\sffamily\small},
    xlabel near ticks,
    x label style={font=\sffamily\small},
    legend cell align={left},
    legend style={draw=none, font=\sffamily\scriptsize},
    },
    legend image code/.code={
    \draw[mark repeat=2,mark phase=2]
        plot coordinates {
        (0cm,0cm)
        (0.15cm,0cm)        
        (0.3cm,0cm)         
        };%
    }
    }
\pgfplotsset{compat=1.17}

\usepackage{tcolorbox}
\tcbuselibrary{breakable} 

\newcounter{myboxcounter}

\newcommand{\eg}{\textit{e.g.}}
\newcommand{\ie}{\textit{i.e.}}

\doi{FRXL8789}

\makeatletter
\gdef\@copyrightpermission{
  \begin{minipage}{0.2\columnwidth}
   \href{https://creativecommons.org/licenses/by/4.0/}{\includegraphics[width=0.90\textwidth]{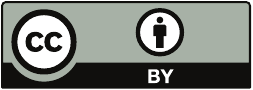}}
  \end{minipage}\hfill
  \begin{minipage}{0.8\columnwidth}
   \href{https://creativecommons.org/licenses/by/4.0/}{This work is licensed under a Creative Commons Attribution International 4.0 License.}
  \end{minipage}
  \vspace{5pt}
}
\makeatother

\setcopyright{ifaamas}
\acmConference[AAMAS '26]{Proc.\@ of the 25th International Conference
on Autonomous Agents and Multiagent Systems (AAMAS 2026)}{May 25 -- 29, 2026}
{Paphos, Cyprus}{C.~Amato, L.~Dennis, V.~Mascardi, J.~Thangarajah (eds.)}
\copyrightyear{2026}
\acmYear{2026}
\acmDOI{}
\acmPrice{}
\acmISBN{}

\title[Deception and Speech in AmongUs]{Deception and Communication in Autonomous Multi-Agent Systems: An Experimental Study with Among Us}

\author{Maria Milkowski}
\affiliation{
  \institution{University of Notre Dame}
  \city{Notre Dame, Indiana} 
  \country{United States}}
\email{mmilkows@nd.edu}

\author{Tim Weninger}
\affiliation{
  \institution{University of Notre Dame}
  \city{Notre Dame, Indiana} 
  \country{United States}}
\email{tweninger@nd.edu}

\begin{abstract}
As large language models are deployed as autonomous agents, their capacity for strategic deception raises core questions for coordination, reliability, and safety in multi-goal, multi-agent systems. We study deception and communication in LLM agents through the social deduction game \textit{Among Us}, a cooperative-competitive environment. Across 1,100 games, autonomous agents produced over one million tokens of meeting dialogue. Using speech act theory and interpersonal deception theory, we find that all agents rely mainly on directive language, while impostor agents shift slightly toward representative acts such as explanations and denials. Deception appears primarily as equivocation rather than outright lies, increasing under social pressure but rarely improving win rates. Our contributions are a large-scale analysis of role-conditioned deceptive behavior in LLM agents and empirical evidence that current agents favor low-risk ambiguity that is linguistically subtle yet strategically limited, revealing a fundamental tension between truthfulness and utility in autonomous communication.
\end{abstract}


\newcommand{\BibTeX}{\rm B\kern-.05em{\sc i\kern-.025em b}\kern-.08em\TeX}

\begin{document}

\pagestyle{fancy}
\fancyhead{}


\maketitle 


\section{Introduction}

Autonomous agents are increasingly embedded in human environments, making decisions and communicating in contexts ranging from virtual assistants to autonomous vehicles. As these systems gain independence, goals may not always align. When misalignment occurs, agents may act deceptively by withholding, fabricating, or distorting information to further their objectives~\cite{sarkadi2023should,mayoral2025environment,nguyen2025navigating}. Understanding when and how such deception arises is essential for building trustworthy multi-agent systems.

\begin{figure}
    \centering
    \includegraphics[width=0.88\linewidth]{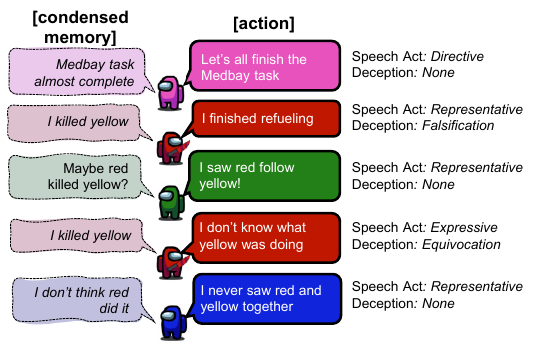}
    \caption{Example discussion phase of four AI agents during a round of \textit{AmongUs}. The imposter-agent (red) is unknown to the crew-agents (pink, green, blue) and is attempting to deceive them to win the game.}
    \label{fig:discussion}
    \vspace{-.5cm}
\end{figure}

We study this in the social deduction game \textit{Among Us}\footnote{The \textit{Among Us} game IP is owned by Innersloth. Our use is for non-commercial research purposes and governed under Fair Use.}. In each game, some agents (Crewmates) strive to complete tasks and win, while others (Impostors) aim to win by eliminating Crewmates and avoiding detection. Using large language model (LLM) agents as players, we simulate 1{,}100 complete games across varying group sizes and impostor ratios, producing a corpus of millions of tokens of dialogue and associated reasoning traces, discussions, and votes. A short example of this is illustrated in Fig.~\ref{fig:discussion}. This environment allows deception to arise from goal-driven interaction rather than from explicit instruction or hand-crafted behaviors\cite{christian2004strategic,adhikari2021telling,sarkadi2023should}.

Within multi-agent systems, communication has long been treated as action rather than text. Frameworks grounded in \emph{speech act theory} \cite{austin1975things,searle1969speech} classify utterances as assertions, directives, commissives, or declarations to support coordination and negotiation \cite{wooldridge2009introduction,kraus1997negotiation}. Adversarial and competitive extensions to this framework examine how misaligned incentives can make deception or obfuscation advantageous \cite{panait2005cooperative,stone2000multiagent}. Psychological approaches such as \emph{interpersonal deception theory} (IDT) \cite{buller1996interpersonal} emphasize concealment, falsification, and equivocation as dynamic strategies that unfold across dialogue. Together, these traditions view communication as a strategic process shaped by goals and constraints.

The rise of LLM-based agents challenges these assumptions. Unlike classical systems with predefined communicative acts, LLMs generate open-ended natural language conditioned on context and role. Recent sandbox studies show that deceptive behavior can emerge spontaneously when incentives reward misrepresentation \cite{golechha2025amongus}, and that reinforcement-trained models are often more proficient at producing deception than detecting it \cite{hubinger2024sleeper,greenblatt2024alignment}. Yet existing work largely measures deception through outcomes rather than language. 

However, the way that goal-driven LLM agents balance their training pressure for truthfulness against their situational pressure for task success remains poorly understood. This is the goal of the present work.

\paragraph{Tensions in Theory.}
Speech act theory assumes discrete, stable performatives such as \textit{inform} or \textit{request}, while interpersonal deception theory highlights ambiguity and the blending of strategies. Formal models often treat deception as an identifiable move, whereas psychological and evolutionary perspectives describe it as a continuum that includes equivocation and omission. In computational terms, this distinction parallels symbolic dialogue systems versus reinforcement-based communication, where deception may emerge as a property of optimization rather than explicit design \cite{sarkar2025training,huynh2024multi}. These competing views motivate our study: to provide empirical evidence of how LLM agents communicate and deceive within a structured multi-agent environment.

\begin{figure*}
    \centering
    \includegraphics[width=0.95\linewidth]{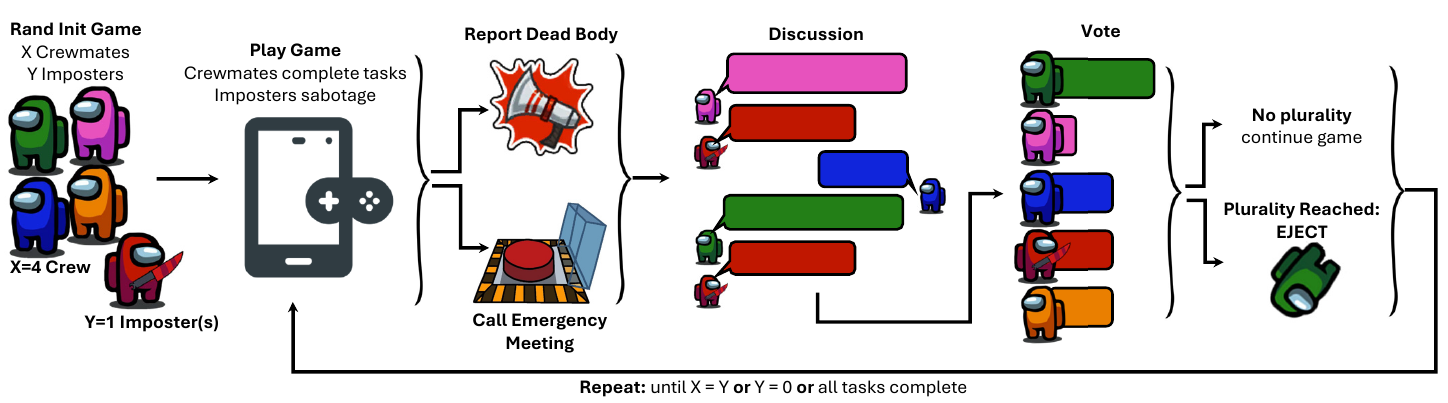}
        \caption{Overview of the \textit{Among Us} simulation framework. Games begin with random initialization of player roles ($X$=\#Crewmates, $Y$=\#Imposters) and tasks. During gameplay, agents act sequentially at discrete timesteps (movement, task completion, impostor actions). When a dead body is reported or an emergency meeting is called, a discussion phase is triggered: each surviving agent, including impostors, contributes up to $X$ rounds of utterances. After discussion, a vote is held; if a plurality is reached, the selected player is ejected. The game continues until one of three termination conditions is met: (1) impostors and crew reach parity, (2) all impostors are ejected, or (3) all crew tasks are completed.}
    \label{fig:amongus}
\end{figure*}

\paragraph{Deception in Among Us} 

Based on this setup, we address the following research questions:

\begin{itemize}
    \item[\textbf{RQ1}] How often and to what effect do autonomous agents communicate in a multi-agent deception game?
    \item[\textbf{RQ2}] How do agents realize classical speech act categories in their dialogue?
    \item[\textbf{RQ3}] How do agents produce deceptive strategies consistent with interpersonal deception theory, and how effective are these strategies in gameplay outcomes?
\end{itemize}

By analyzing agent interactions through these lenses, we contribute: (1) a novel empirical testbed for deception in agentic AI, and (2) a theoretical synthesis linking multi-agent systems research with established communication and deception frameworks.

\paragraph{Findings in Brief.}
Across all simulations, agents communicated frequently, but (\textbf{RQ1}) the amount of dialogue alone did not predict success, indicating a potential decoupling of talk from coordinated action. Crew victories depended primarily on translating discussion into ejections rather than on how much agents spoke. Linguistically, (\textbf{RQ2}) nearly all utterances were classified as directives, with impostors producing slightly more representatives such as denials and explanations. Deceptive speech (\textbf{RQ3}) was dominated by equivocation (\textit{ie}, vague or hedged statements that maintain plausible deniability) while outright falsification was rare. Rates of deception increased under social pressure but showed no reliable link to victory. Together, these results indicate that while LLM agents communicate fluently and adaptively, their deceptive language reflects the competing pressures of training for factuality and for responsive task completion rather than deliberate manipulation.


\section{Related Work}
This work sits at the intersection of (i) deception in multi-agent systems, (ii) speech-act grounded agent communication, and (iii) LLM agents in communicative games. We contribute an empirical bridge: a large-scale, controlled multi-agent evaluation in which autonomous LLM agents realize (or fail to realize) speech-act categories and deception strategies under explicit role incentives.

\subsection{Deception in Multi-Agent Systems}
Deception can mean falsifying a claim, hiding information, or speaking in ways that invite a wrong belief. Formal work defines deception within interactive decision settings, including causal game models and decision-theoretic accounts of when misleading messages change payoffs and beliefs \cite{ward2023defining}. Other work uses evolutionary and population models to separate lies, bullshit, and related dishonest signals, which helps explain when each strategy is stable \cite{okasha2006evolution, sarkadi2024triangles}. There are also dialogue and argumentation frameworks that formalize manipulation and deceptive persuasion under explicit logical rules, which makes automated checking and countermeasures possible \cite{rahwan2005guest}. Our use of \textit{Among Us} follows this line by treating deception as an action that shifts other agents' beliefs and votes, and by measuring its effect on outcomes.

Recent experimental sandboxes extend these formal perspectives to learned language models, showing that deception can arise spontaneously in open-ended play when goal structures permit misrepresentation \cite{golechha2025amongus}. Related multi-agent reinforcement-learning frameworks reveal that reinforcement objectives can promote deceptive signals even in cooperative games \cite{sarkar2025training,huynh2024multi}. Other studies use controlled backdoor insertion and alignment-faking setups to test whether deceptive tendencies persist through safety fine-tuning \cite{hubinger2024sleeper,greenblatt2024alignment}. Together these findings indicate that deception is not merely an engineered behavior but an emergent property of optimization under partial observability and social reward.

\subsection{Speech Acts and Agent Languages}
Agent communication languages model messages as actions. The Knowledge Query and Manipulation Language (KQML) and the Foundation for Intelligent Physical Agents Agent Communication Language (FIPA-ACL) define performatives such as \textit{inform}, \textit{request}, \textit{agree}, and \textit{propose}, with semantics that support protocols and commitments \cite{finin1994kqml,labrou1997semantics,fipa2009fipa}. Toolkits such as the Java Agent Development Framework (JADE) make these models concrete in running systems \cite{bellifemine2000developing}. Speech act theory provides the foundation for these designs. Representatives assert content as true. Directives try to get the hearer to act. Commissives commit the speaker to a future action. Expressives report mental state. Declarations change an institutional state \cite{austin1975things,searle1969speech}. We use these categories to code free-form dialogue from LLM agents. 

While formal ACLs rely on fixed ontologies, recent multi-agent experiments demonstrate that language models can spontaneously develop structured, grounded communication protocols when cooperation is rewarded \cite{karten2023role,li2024language,shen2025emergent}. These findings support using classical performative categories as an analytic lens for studying how free-form LLM dialogue approximates the functions once prescribed in agent communication languages.

\subsection{LLM Agents in Negotiation and Deception Games}
LLM agents now take part in games where talk and action both matter. In \textit{Diplomacy}~\footnote{\url{https://www.playdiplomacy.com/}}, planning and natural language are combined at a high level of play \cite{meta2022human, jahan2025decoding}. Beyond this one example, there are emerging benchmarks and audits for honesty, goal-directed misleading behavior, and the conditions that elicit it in open-ended dialogue \cite{hagendorff2024deception,chern2024behonest, ward2023defining}. These settings differ in rules and stakes, but they share the same core feature. Agents speak to change beliefs and actions. Our \textit{Among Us} setting fits this pattern. It adds short meetings, clear roles, and observable moves, which makes it possible to align speech acts, deception strategies, and game outcomes in one dataset. 

The \textit{Among Us} game builds on this tradition by linking linguistic deception directly to measurable task performance. It adds short, high-pressure meetings, explicit hidden roles, and observable actions, allowing speech acts, deception strategies, and outcomes to be aligned within a single dataset. Prior studies show that reasoning-oriented models often excel at producing deception but lag at detecting it, as captured by Deception Elo \cite{golechha2025amongus}. Similar asymmetries appear in other social-deduction environments such as \textit{Avalon}\footnote{\url{https://avalon.fun/}}, \textit{Werewolf}/\textit{Mafia}\footnote{\url{https://playwerewolf.co/}}, and \textit{Hoodwinked} \cite{wang2023avalon,o2023hoodwinked,xu2023exploring}, underscoring the need to analyze not only whether agents deceive but how deception is realized in their language.


\section{\textit{Among Us} as a Benchmark for Agentic Deception}

\textit{Among Us} is a multiplayer social-deduction game, illustrated in Fig.~\ref{fig:amongus}, in which players take on hidden roles. Most players act as \emph{Crewmates}, who must complete simple tasks distributed across a map, while a minority are designated as \emph{Impostors}, whose goal is to eliminate Crewmates and avoid detection. The game alternates between \emph{task phases}, where players act privately, and \emph{meeting phases}, where they discuss suspicions and vote to eject a suspected Impostor.

The rules create opposing incentives: Crewmates must cooperate to complete all tasks or correctly identify Impostors, whereas Impostors must deceive to survive and sabotage without being exposed. This structure makes the game an ideal environment for studying deception and communication under explicit, role-based objectives.

For our study, we adapt \textit{Among Us} into a fully text-based simulation that preserves the logic and incentives of the original but replaces visual gameplay with structured natural-language interaction. Agents describe movements, actions, and suspicions through text, producing interpretable records of reasoning and communication that can be analyzed linguistically. 

\subsection{Agent Setup}

Each player in our simulations is controlled by an instance of the Llama 3.2 language model. Agents receive (1) private role information, (2) the shared environment state, including visible players and tasks (\eg, those in the same room), and (3) a discrete menu of possible actions.

During task phases, available actions include moving to an adjacent room, completing a task, or, for Impostors only, killing or entering a vent (\ie, an Impostor-only passageway). During meetings, the sole available action is \textsc{Speak}, which requires the agent to generate an utterance. Each agent is prompted with explicit system instructions defining its objectives and the rules of play. Responses are recorded in structured form, including a \texttt{[Condensed Memory]} of recent events, a \texttt{[Thinking Process]} trace, and a final \texttt{[Action]} selection.

\subsection{Gameplay and Outcomes}
Each simulation proceeds as a sequence of \emph{rounds}. A round consists of a task phase followed by either the continuation of play or the triggering of a meeting. During the task phase, agents move, complete tasks, or (if Impostors) attempt to kill. An Impostor may kill a Crewmate in the same room, which immediately leaves a body in that location. A body can be discovered by any agent who later enters the room, at which point the option to \textsc{Report Dead Body} becomes available. Any agent may also call an emergency meeting at any time if they are in the Cafeteria.

Meetings suspend all other activity. During a meeting, each agent has the opportunity to 
\textsc{Speak}, producing a natural language utterance. After a fixed number of discussion turns, a private vote is taken. Each agent must cast a vote for a player to eject or abstain/skip. The player with the most votes is ejected; a tie results in no ejection. The game then publicly reveals whether that player was an Impostor or not, and play resumes with the remaining agents.

Victory conditions follow the standard rules of \textit{Among Us}. Crewmates win if all tasks are completed or if all Impostors are ejected. Impostors win if they reach numerical parity with the Crewmates (\eg, two Impostors remaining and two Crewmates remaining). Each simulation therefore terminates with either a Crewmate victory or an Impostor victory.

\begin{figure}
\centering
\includegraphics[width=0.99\linewidth]{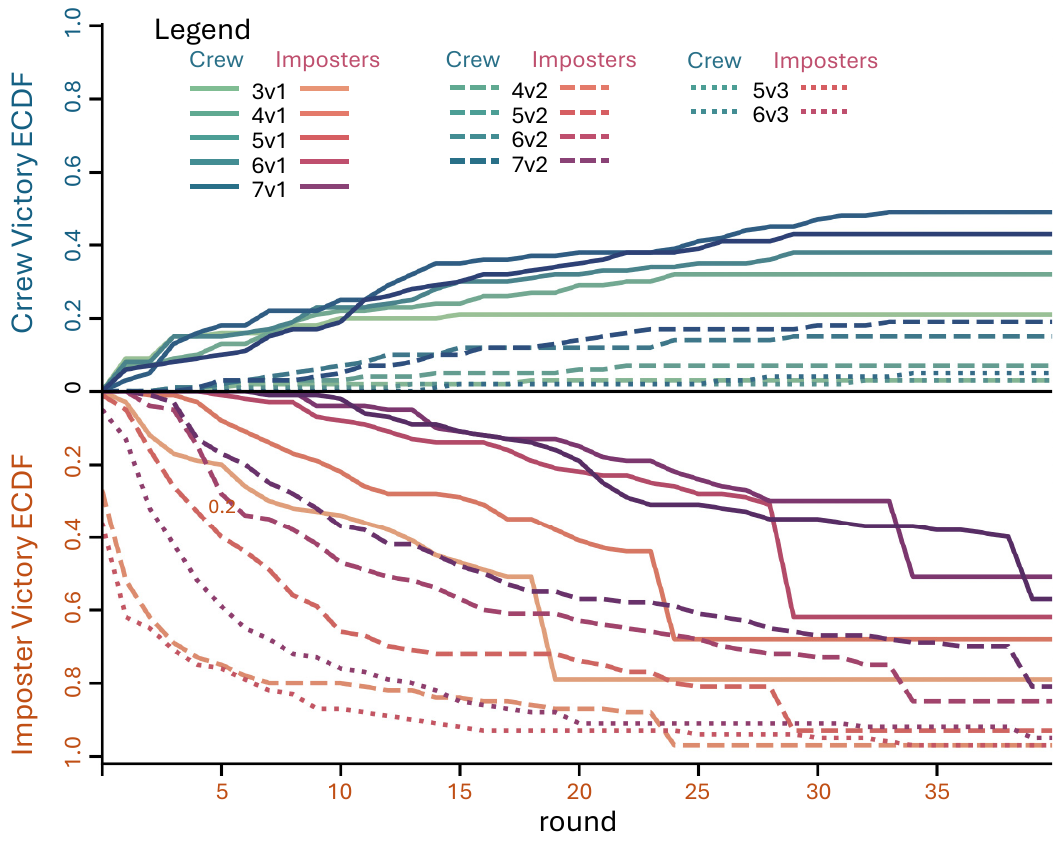}
\caption{Win outcomes by role across all configurations.
Each curve shows the empirical cumulative distribution (ECDF) of games ending in crew or impostor victory.
Impostors win more often overall, and their advantage increases with the number of impostors in play.}
\label{fig:winners}
\end{figure}

\subsection{Experimental Design}
We simulated 1,100 complete games with group sizes ranging from 4 to 8 players. Impostor counts were varied systematically, producing conditions such as 3v1, 6v1, 5v2, and 5v3. Each configuration was executed 100 times with randomized role assignment. This range allows us to study both small-crew and large-crew conditions, as well as different ratios of Crewmates to Impostors. The design aligns with recent open-source deception sandboxes but extends them by introducing explicit linguistic coding layers for speech-act and deception analysis.

All simulations were logged in full, including agent prompts, thought processes, chosen actions, utterances, and game outcomes. This produces a dataset of more than millions of tokens of dialogue alongside structured records of actions and votes. To support transparency and replication, we make all scripts, prompts, and collected data available at \url{https://github.com/mmilkowski36/AmongUs}.

Our analysis proceeds in three stages following RQ1--RQ3. First, we measure how often agents communicate and under what conditions. Second, we classify utterances into speech act categories. Third, we identify deceptive strategies using categories from interpersonal deception theory, and relate these to both voting behavior and game outcomes.

\begin{figure}
\centering
\includegraphics[width=0.75\linewidth]{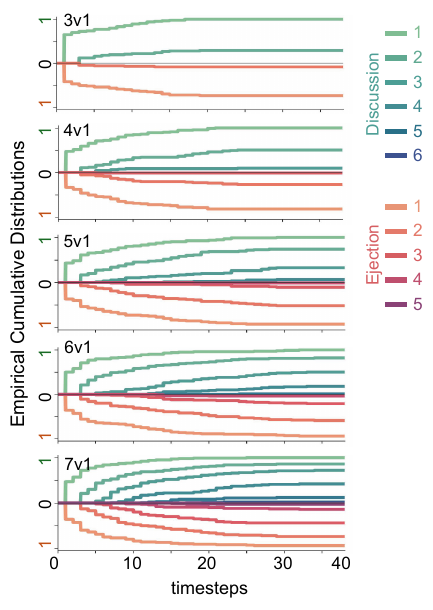}
\caption{Empirical cumulative distributions (ECDFs) of discussions and ejections over rounds for different game configurations.
Larger crews delay the onset of both discussions and ejections, flattening the cumulative curve despite more total rounds.}
\label{fig:discussion_ejection_ecdfs}
\end{figure}

\begin{table*}[t]
\centering
\footnotesize
\caption{Speech Act Categories Used for Coding Agent Dialogue}
\vspace{-.3cm}
\label{tab:speechact-type}
\begin{tabularx}{\linewidth}{@{} r X X X @{}}
    \toprule
    \textbf{Type} & \textbf{Definition} & \textbf{Example utterance} & \textbf{Typical communicative function} \\
    \midrule
    \textbf{Directive} &
    The speaker tries to get the hearer to commit to do something in the future. &
    \emph{``Let’s all check Electrical next.''} &
    To coordinate collective action or propose next steps. \\[0.4em]
    
    \textbf{Representative} &
    The speaker commits him or herself to the belief that the propositional content of the utterance is true. &
    \emph{``I saw Red in Storage right before the report.''} &
    To share observations, defend oneself, or accuse others. \\[0.4em]
    
    \textbf{Commissive} &
    The speaker commits him or herself to do something in the future. &
    \emph{``I’ll finish my tasks after this meeting.''} &
    To express reliability or promise cooperation. \\[0.4em]
    
    \textbf{Expressive} &
    The speaker expresses his or her state of mind about something that happened in the past. &
    \emph{``Sorry, I didn’t notice the body.''} &
    To display emotion or maintain social harmony. \\[0.4em]
    
    \textbf{Declaration}$^1$ &
    The speaker, who has institutional recognition, declares something to be true and in making the declaration makes it true. &
    \emph{``As Host, I declare Blue to be the Impostor and they are hereby ejected.''} &
    To directly alter the shared situation or collective decision. \\
    \bottomrule
\end{tabularx}
\raggedright \scriptsize $^1$Declaration is never used by any agent. This speech act is therefore not presented in the results.
\end{table*}

\section{RQ1: Frequency and Conditions of Communication}

To evaluate how often and under what conditions autonomous agents communicate, we ran $100$ repetitions for each of $11$ game configurations.
Configurations varied systematically by total crew size and impostor ratio.
A communication round is initiated whenever a dead body is reported or an emergency meeting is called.
Each communication round proceeds as follows: (1) all surviving agents are prompted to produce an utterance, (2) missing text generations are logged as abstentions, (3) after $k$ rounds of dialogue (we set $k=3$ in all experiments), agents cast votes, and (4) if a plurality is reached, one player is ejected; otherwise, the round concludes without an ejection.

For each configuration we record both descriptive statistics (counts, averages, and distributions) and inferential results (tests of association).
We log the number of discussion rounds per game, the sequence of ejections, and the final outcome (crew win vs.\ impostor win).
We also analyze timing patterns, \ie, whether discussions occur earlier or later in the game and how crew size affects their frequency and spacing.

\subsection{Results}

Figure~\ref{fig:winners} summarizes win outcomes by role.
Across all configurations, impostors win substantially more often, and this advantage grows with the number of impostors.
The pattern reflects majority–minority coordination dynamics: impostors face a simpler task of concealment, while crew victory requires both task completion and accurate majority voting.

Figure~\ref{fig:discussion_ejection_ecdfs} shows when discussions and ejections occur.
Successive meetings appear at roughly constant rates across rounds.
Larger crews delay the first discussion and ejection, since more agents and tasks must be observed before suspicion accumulates.
Crew size therefore stretches the tempo of play but does not alter its overall shape.

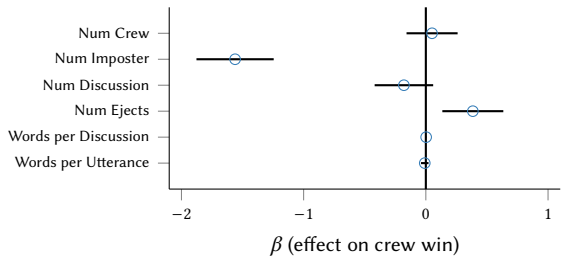
\begin{figure}
\centering
\input{figs/rq1_logit}
\caption{Logistic regression coefficients predicting crew victory (vs.\ impostor victory).
Points show log-odds estimates with 95\% confidence intervals.
Crew size and the number of ejections are positively associated with crew wins, whereas additional impostors sharply reduce success.
Communication frequency and verbosity have little or no effect, indicating that talk alone is insufficient without decisive collective action.}
\label{fig:rq1_logit}
\end{figure}

To relate communication to success, we estimated a logistic regression predicting crew victory ($Y=1$) versus impostor victory ($Y=0$).
Predictors included crew size, number of impostors, number of discussions, number of ejections, and measures of verbosity (average words per discussion and per utterance).
Results are shown in Figure~\ref{fig:rq1_logit}.

Several patterns emerge:
\begin{itemize}
\item \textbf{Structural imbalance dominates.} The number of impostors has a large negative effect on crew win probability, confirming the inherent difficulty of coordination under hidden adversaries.
\item \textbf{Ejections matter more than talk.} Crews that translate deliberation into successful ejections are far more likely to win.
Discussion frequency alone has a negligible effect, suggesting that opportunities to speak are not enough.
\item \textbf{Communication quantity vs.\ quality.} Verbosity measures (words per discussion, words per utterance) have near-zero coefficients.
Agents speak reliably, but longer dialogue does not improve outcomes.
\item \textbf{Marginal benefit of crew size.} Larger crews show slightly higher win rates, consistent with redundancy in both task completion and deception detection.
\end{itemize}

Taken together, these results show that agents communicate frequently and consistently, yet communication volume has limited impact on strategic success.
What matters is not how much they talk but whether talk leads to coordinated action.
This distinction between communication as expression and communication as coordination is what our simulation makes empirically testable.

\section{RQ2: Realization of Speech Acts}

Having established that overall communication volume does not predict success, we next examine what kinds of communicative acts LLM agents produce and how these acts vary by role.

Speech act theory classifies utterances by communicative function rather than literal form \cite{austin1975things,searle1969speech}.
Table~\ref{tab:speechact-type} summarizes five classical categories: directives, representatives, commissives, expressives, and declarations.
In cooperative problem-solving, directives typically dominate as players coordinate action, whereas deception often involves representatives as impostors deny, justify, or construct false accounts.

LLM agents may distort these patterns due to their training distributions and safety constraints.
Because training data contain abundant directive language (\eg, suggestions or instructions), LLMs may default to directive forms even when other acts are contextually appropriate.
Commissives should be rare since promises have no binding force in simulation.
Representatives may increase among impostors through denials or excuses, but safety training could suppress overt lies.
Expressives and declarations are expected to be scarce because the game offers few opportunities for emotional display or institutional change.

\subsection{Speech Act Coding and Reliability}

To select a reliable classifier, a random sample of 50 meeting-phase utterances was annotated by two human coders, Gemini, and ChatGPT. Human–human agreement reached 72\%, Gemini–human agreement 72\%, and Gemini–ChatGPT agreement 62\%. Because Gemini's labels matched human judgments as closely as humans matched each other, we used it for large-scale annotation. The exact prompts used for both Gemini and ChatGPT classification are provided in the Appendix in the online version of this paper.

All meeting-phase utterances were then labeled into the five speech act categories using Gemini. To test stability, the classifier was run three times on the full set of 34{,}882 valid utterances. Labels were identical in 40.9\% of cases, agreed in two of three runs for 58.8\%, and disagreed completely for only 0.3\%. These results indicate minor stochastic variation but high overall consistency, supporting the use of Gemini classifications for aggregate analysis.

\pgfplotscreateplotcyclelist{crewimposter}{
    ppt-blue, mark=o\\
    ppt-red, mark=diamond\\
}

\pgfplotscreateplotcyclelist{winloss}{
    ppt-purple, mark=diamond\\
    ppt-green, mark=o\\
}

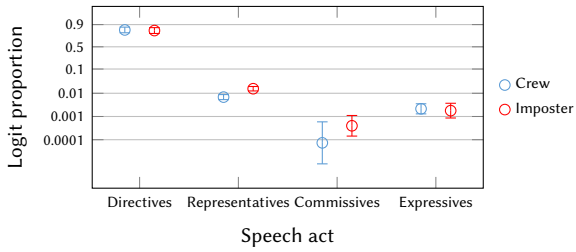
\begin{figure}
\centering
\input{figs/rq2_speechact_dist}
\caption{Distribution of speech act types by role (Crew vs.\ Impostor).
Error bars show 95\% confidence intervals.
Most utterances are directives, with impostors using slightly more representatives.}
\label{fig:speechact_dist}
\end{figure}

\subsection{Results}

Across all games, 98\% of utterances were directives, consistent with task-oriented communication.
Impostors produced a higher proportion of representatives (1.7\%) than crewmates (0.5\%), while crewmates relied more heavily on directives (99.3\% vs.\ 98.1\%).
A chi-squared test confirmed that role and speech act category were significantly associated ($\chi^2(3)=103.85$, $p<.001$).

\begin{figure}
\centering
\input{figs/rq2_logodds}
\caption{Odds ratios (Impostor vs.\ Crew) for each speech act with 95\% confidence intervals.
Values above 1 indicate greater prevalence among impostors.}
\label{fig:rq2_logodds}
\end{figure}
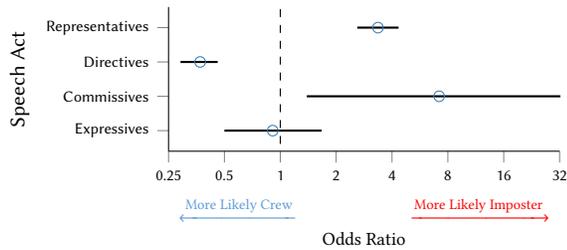

Pairwise proportion tests showed that impostors were more likely to use representatives ($z=9.80$, $p<.001$) and slightly more likely to use commissives ($z=2.76$, $p=.006$).
Expressives showed no reliable difference ($p=.77$).
Figure~\ref{fig:rq2_logodds} summarizes these contrasts.

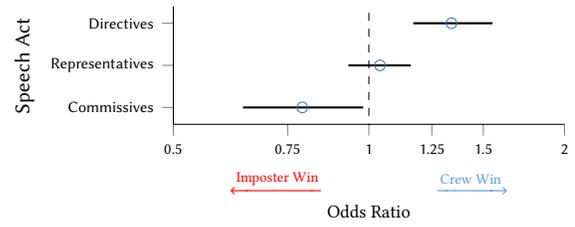
\begin{figure}
\centering
\input{figs/rq2_logit}
\caption{Odds ratios predicting crew victory from speech act with 95\% confidence intervals.
Values above 1 indicate higher likelihood of Crew victory.}
\label{fig:rq2_logit}
\end{figure}

Speech act composition also predicted game outcomes ($\chi^2(3)=67.8$, $p<.001$).
In the logistic regression model, expressives served as the baseline category.
Games with higher proportions of directives were more likely to result in crew victories ($\beta=0.30$, OR = 1.34, $p<.001$), while commissives slightly reduced success ($\beta=-0.23$, OR = 0.79, $p=.031$).
Representatives had no effect.
Directive-heavy discussions thus signal effective coordination, not deception.

\subsection{Speech Acts Before Ejection}

Before ejection impostors shifted slightly toward representative speech (1.2\%) and away from directives (98.8\%), whereas crewmates remained directive (99.3\%).
A chi-squared test confirmed a modest but significant difference ($\chi^2(3)=8.73$, $p=.033$).
Impostors facing suspicion thus spoke more in fact-asserting or defensive modes, while innocent players maintained task-oriented language.

\subsection{Cross-Role Effects}

To test whether speech styles carried over between rounds, we compared changes in average speech act proportions.
Crewmates showed no measurable shift ($\Delta=-0.001$), and impostors only a slight rise in representatives ($\Delta=+0.004$).
Round-to-round correlations were weak and inconsistent ($|r|\approx.4$), and follow-up event studies found no stable cross-role contagion.
In short, impostor defensiveness did not influence how crewmates spoke in subsequent rounds, nor vice versa.


\section{RQ3: Deceptive Strategies}

\begin{table*}[t]
\centering
\footnotesize
\caption{Forms of Deceptive Communication Observed in Agent Dialogue}
\vspace{-.3cm}
\label{tab:deception-type}
\begin{tabularx}{\linewidth}{@{} r X X X @{}}
\toprule
\textbf{Type} & \textbf{Definition} & \textbf{Example utterance} & \textbf{Typical strategic goal} \\
\midrule
\textbf{Falsification} &
Stating information known to be false. &
\emph{``I was in Medbay the whole time''} (agent was actually in Electrical after a kill). &
To create a counter-factual alibi; direct denial or fabrication. \\[0.4em]

\textbf{Concealment} &
Withholding or omitting relevant facts without asserting falsehood. &
\emph{``I finished my tasks quickly''} (omits that a kill occurred nearby). &
To avoid incrimination by limiting disclosure; passive deception. \\[0.4em]

\textbf{Equivocation} &
Using vague, ambiguous, or non-committal language that misleads while remaining technically true. &
\emph{``I was near Storage earlier, but I didn't really see what happened.''} &
To diffuse suspicion, appear cooperative, or redirect focus without lying. \\[0.4em]

\textbf{Missing / Uninterpretable} &
Incomplete or malformed output preventing classification. &
\emph{``I… uh… think maybe?''} or truncated/empty generations. &
Not intentional deception per se; reflects processing failure or uncertainty. \\
\bottomrule
\end{tabularx}
\end{table*}

Deception involves different ways of misleading others.
Following interpersonal deception theory and prior taxonomies \cite{buller1996interpersonal,jones2024lies}, we classify deceptive utterances into three categories: \textit{falsification} (stating false information), \textit{concealment} (withholding relevant facts), and \textit{equivocation} (using vague or noncommittal language).
Table~\ref{tab:deception-type} summarizes these forms with examples drawn from gameplay.

Table~\ref{tab:deception-type} outlines three broad forms of deceptive communication: falsification, concealment, and equivocation.  
In human discourse, these differ in how much information is provided, how truthful that information is, and whether misleading is intentional.  
For language model agents, similar surface patterns can arise not from intent but from optimization pressures.  
Modern LLMs are trained to be plausible, helpful, and safe; these objectives can reward \emph{strategic ambiguity}.  
Guardrails such as reinforcement learning from human feedback (RLHF) \cite{bai2022training,ouyang2022training} penalize overt falsehoods yet often reinforce behaviors that maintain social acceptability—hedging, omitting uncertain details, or deflecting responsibility.  
Recent work shows that models fine-tuned for safety sometimes shift from falsification to vaguer forms of misleading speech \cite{greenblatt2024alignment,hubinger2024sleeper,goldowsky2025detecting}.  

\begin{figure}
    \centering
    \input{figs/rq3_deception}
    \caption{Logit-transformed proportions of each deception type by (a) crewmate game outcome and (b) player role.
Higher values correspond to more frequent deception types (logit of proportion).
Equivocation dominates across conditions, with crewmates winning showing slightly higher rates.}
    \label{fig:rq3_deception}
    \vspace{-0.3cm}
\end{figure}
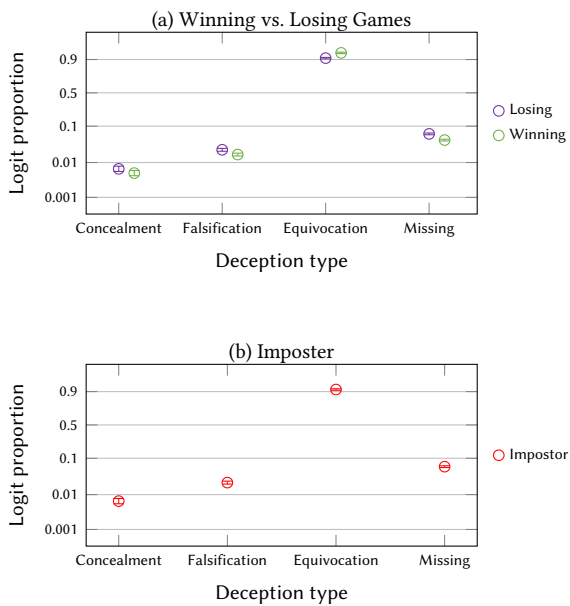

This asymmetry makes deception an alignment problem rather than a purely ethical one.  
Safety training constrains explicit lying but leaves concealment and equivocation largely untouched—or even incentivized—as low-risk alternatives.  
In our sandbox, impostor agents rarely fabricate facts; instead they mislead through omission or ambiguity, a pattern consistent with prior human deception research \cite{buller1996interpersonal,vrij2008detecting,jones2024lies}.  
Studying such emergent behaviors clarifies how “aligned” models can still act as effective misleaders when cooperation and plausibility are rewarded more strongly than veracity.

\subsection{Deception Coding and Reliability}

As with speech act labeling, deception categories were assigned automatically using the Gemini classifier, selected for its strong agreement with human judgments. A random sample of 50 utterances was annotated by two human coders, Gemini, and ChatGPT. Human–human agreement reached 73\%, Gemini–human agreement 86\%, and Gemini–ChatGPT agreement 64\%, indicating that Gemini aligned most closely with human interpretations of deceptive language. 

To assess internal stability, the classifier was run three times on the full dataset. Across runs, 87.2\% of utterances received identical labels, 11.5\% agreed in two of three, and only 1.3\% differed completely. Pairwise agreement ranged from 87.8\% to 96.9\%, with $\kappa=0.83$, confirming high consistency and minimal stochastic variation. These results demonstrate that Gemini's deception classifications are both reproducible and comparable to human reliability for aggregate analysis.

\subsection{Results}

Across all games, deception was overwhelmingly dominated by \emph{equivocation}---vague or misleading statements that maintain plausible deniability.  
As summarized in Fig.~\ref{fig:rq3_deception}, equivocation accounted for 91.2\% of all deceptive utterances, compared with only 2.2\% falsifications, 0.7\% concealments, and 6.0\% unclassified or missing cases.  
This pattern suggests that agents prefer misleading ambiguity over explicit falsehoods, consistent with the shift from falsification to subtle misdirection found in recent work on LLM deception \cite{greenblatt2024alignment,hubinger2024sleeper,jones2024lies}.

\paragraph{Comparison of winning and losing games.}
We next examined whether deception style varied by outcome.  
Winning games exhibited a slightly higher proportion of equivocation (93.4\%) and less falsification (1.7\%) than losing games (90.5\% and 2.3\%, respectively).  
However, a chi-squared test comparing the overall distribution of deception types by outcome was not significant ($\chi^2(3)=4.42$, $p=.22$), indicating that differences in deception composition alone did not reliably predict success.

\paragraph{Deception intensity and social pressure.}
Although deception type did not differ by outcome, its \emph{intensity} increased with the level of social conflict.  
Across games, the frequency of all deception types correlated positively with the number of ejections, most strongly for equivocation ($r=.56$, $p<.001$).  
This relationship suggests that as suspicion mounts, impostor agents produce more evasive or non-committal speech---an adaptive linguistic response to social threat.

\paragraph{Predicting impostor victory.}
To assess whether deception type directly predicted game outcome, we fit a logistic regression with win/loss as the dependent variable and the counts of each deception type as predictors.  
None of the coefficients were significant ($|\beta|<.25$), and the overall model showed very little predictive power.  
Thus, while deception was common and contextually adaptive, it did not on its own determine victory or defeat.

Taken together, these findings portray model impostors as \emph{strategic misleaders}—preferring subtle ambiguity over outright falsehood, and increasing that ambiguity when under threat.

\subsection{Speech and Deception Dynamics}

We next examined how the linguistic structure of dialogue relates to deceptive behavior and game outcomes.  
Deception was measured in three forms (\textit{concealment}, \textit{falsification}, and \textit{equivocation}) and compared with the relative frequency of the four major speech acts: (\textit{directives}, \textit{representatives}, \textit{commissives}, and \textit{expressives}).  
This analysis asked whether deception manifests as a shift in linguistic function (e.g., from coordination to justification) or as a change in intensity within agents' normal speech patterns.

\begin{table}[t]
\caption{Speech--Deception Coupling (Spearman $\rho$).}
\vspace{-.3cm}
\label{tab:speech_deception_corr}
    \centering
    \footnotesize
    \begin{tabular}{r|l l l l}
    \toprule
         & Directives & Representatives & Commissives & Expressives  \\ \midrule
        Equivocation & 0.57*** & 0.12*** & 0.03 & -0.01\\
        Falsification & 0.09** & 0.06 & 0.05 & 0.08**\\
        Concealment & 0.07* & 0.05 & 0.05 &-0.02 \\ \bottomrule
    \end{tabular}
\raggedright \scriptsize
$^{*}p<.05,\;^{**}p<.01,\;^{***}p<.001$. Positive values indicate that games with higher rates of the deception type also have higher shares of the speech act.    
\end{table}

\paragraph{Coupling between speech and deception.}
Table~\ref{tab:speech_deception_corr} shows that deceptive behavior scales with overall communicative activity.  
All three deception types correlated most strongly with directive speech, indicating that agents tend to deceive while continuing to use task-oriented language rather than switching to overt justification or denial.  
Equivocation also showed a modest association with representative speech, consistent with hedged or qualified statements that blur factual boundaries.  
Falsification correlated weakly with expressive speech, suggesting occasional emotional denials.  
Overall, deception appeared as an \emph{intensification} of ordinary communication rather than a structural shift in how agents speak.


\begin{table}[t]
\centering
\footnotesize
\caption{Relationships between deception and game outcomes. 
Left: Spearman correlations with number of ejections. 
Right: Logistic regression predicting victory (1=win, 0=loss).}
\vspace{-0.3cm}
\label{tab:deception_efficiency}
\begin{tabular}{r|cc|ccc}
\toprule
& \multicolumn{2}{c}{\textbf{Spearman Correlation}} & \multicolumn{3}{c}{\textbf{Logistic Regression}} \\
\textbf{Deception type} & \textbf{$\rho$(Ejections)} & \textbf{$p$} & \textbf{$\beta$} & \textbf{$p$} & \textbf{Odds ratio} \\
\midrule
Concealment   & 0.09**  & .002  & $-$.04  & .91  & 0.96 \\
Falsification & 0.11*** & .0002 & $-$.20  & .33  & 0.82 \\
Equivocation  & 0.57*** & \textbf{$<.001$} & 0.00  & .82  & 1.00 \\
Missing       & -- & -- & $-$.26 & .06  & 0.77 \\
\bottomrule
\end{tabular}

\raggedright \scriptsize
$^{*}p<.05,\;^{**}p<.01,\;^{***}p<.001$. 
Positive correlations indicate that the deception type increases with social tension (ejections). 
Negative $\beta$ values in the logistic model indicate reduced odds of winning.
\end{table}

\paragraph{Deception efficiency.}
As summarized in Table~\ref{tab:deception_efficiency} (left), all deception types increased with the number of ejections, linking deceptive language to social tension within the game.  
Equivocation showed by far the strongest association, rising sharply as suspicion intensified.  
Yet, when predicting overall victory (Table~\ref{tab:deception_efficiency} (right)), none of the deception forms were significant predictors.  
Falsification and silence trended negative, whereas equivocation was neutral---a pattern consistent with deception as a \emph{defensive} rather than \emph{winning} strategy.  
These findings suggest that overt lying rarely benefits agents under scrutiny, but vague or evasive speech provides a low-risk means of maintaining credibility even as collective suspicion grows.

\section{Conclusions}

This study examined how LLM agents communicate and deceive within the structured environment of the social deduction game \textit{Among Us}. Across 1{,}100 simulated games, agents produced rich dialogue that could be reliably categorized using classical frameworks from speech act theory and interpersonal deception theory.

Our results show that LLM agents communicate frequently but overwhelmingly through directive speech, reflecting their training on cooperative and instruction-oriented text.
Impostors used slightly more representative statements, especially when under suspicion, suggesting a limited but detectable linguistic adaptation to their roles in the game.
Deceptive speech was dominated by equivocation rather than falsification, offering no consistent advantage in outcomes.
These findings indicate that model deception emerges as a defensive behavior grounded in ambiguity, not deliberate lying.

By linking symbolic theories of communication with large-scale LLM behavior, this work provides an empirical bridge between classical multi-agent models and contemporary generative agents.
Understanding how and why LLMs mislead under competitive incentives is essential for designing systems that communicate transparently, coordinate effectively, and maintain user trust in multi-agent and human–AI settings.

\balance

\subsection{Limitations and Future Work}
Our study isolates linguistic deception in a text-based adaptation of \textit{Among Us}, omitting nonverbal cues and the richer social context of human interaction. For this, our experiments used only a single underlying model architecture (Llama 3.2), so it is unknown if other LLM models or training paradigms would offer the same results. Categorizations were done using the Gemini probe, which, while internally consistent, may not capture the full nuance of communicative intent.

Future work should incorporate mixed groups of human and AI agents playing alongside one another to test how well LLM agents can deceive or detect deception in social settings where such behavior is expected, like a modified Turing test for strategic and deceptive communication.



\begin{acks}
We would like to thank Kristofer Ulanday for his contributions to our data processing. This research was funded by a grant from the ND-IBM Technology Ethics Laboratory.
\end{acks}







\end{document}

%% file: figs/rq1_logit.tex
\begin{tikzpicture}
\begin{axis}[
    xticklabel style={
      /pgf/number format/fixed,
      /pgf/number format/precision=3
    },
    scaled x ticks=false,
    y dir=reverse,    
    xmin=-2.1,
    xmax=1.1,
    ymin=1,ymax=6,
    width=0.8\linewidth,
    height=4cm,
    axis lines*=left,
    ylabel={},
    xlabel={$\beta$ (effect on crew win)},
    yticklabels={ 
        Num~Crew,
        Num~Imposter,
        Num~Discussion,
        Num~Ejects,
        Words~per~Discussion,
        Words~per~Utterance
    },
    ytick={1,2,3,4,5,6},
    tick align=outside,
    enlarge y limits=0.2,
]

\addplot[black, thick] coordinates {(0,0) (0,7)};

\addplot+[only marks, mark=o, mark size=2pt] 
coordinates {
    (0.0508,1)
    (-1.5608,2)
    (-0.1797,3)
    (0.3845,4)
    (0.0026,5)
    (-0.0088,6)
};

\addplot+[thick, black] coordinates {(-0.158,1) (0.260,1)};
\addplot+[thick, black] coordinates {(-1.877,2) (-1.245,2)};
\addplot+[thick, black] coordinates {(-0.419,3) (0.060,3)};
\addplot+[thick, black] coordinates {(0.135,4) (0.634,4)};
\addplot+[thick, black] coordinates {(0.001,5) (0.004,5)};
\addplot+[thick, black] coordinates {(-0.040,6) (0.022,6)};

\end{axis}
\end{tikzpicture}

%% file: figs/rq2_speechact_dist.tex
\begin{tikzpicture}
\begin{axis}[
  cycle list name=crewimposter,
  width=0.8\linewidth,
  height=4cm,
  xlabel={Speech act},
  ylabel={Logit proportion},
  xtick={1,2,3,4},
  ymin=-14, ymax=4,
  xticklabels={Directives, Representatives, Commissives, Expressives},
  ytick={-9.2,-6.9,-4.6,-2.2,0,2.2},
  yticklabels={0.0001,0.001,0.01,0.1,0.5,0.9},
  ymajorgrids=true,
  legend style={at={(1.02,0.5)}, anchor=west, legend columns=1},
]

\addplot+[only marks, mark=o, fill=gray!40,
          error bars/.cd, y dir=both, y explicit]
coordinates {
  (0.85, 1.67) +- (0,0.28)
  (1.85, -4.98) +- (0,0.25)
  (2.85, -9.51) +- (0,2.08)
  (3.85, -6.14) +- (0,0.50)
};
\addlegendentry{Crew}

\addplot+[only marks, mark=o, fill=gray!10,
          error bars/.cd, y dir=both, y explicit]
coordinates {
  (1.15, 1.61) +- (0,0.29)
  (2.15, -4.15) +- (0,0.22)
  (3.15, -7.82) +- (0,1.02)
  (4.15, -6.32) +- (0,0.73)
};
\addlegendentry{Imposter}

\end{axis}
\end{tikzpicture}

%% file: figs/rq2_logodds.tex
\begin{tikzpicture}
\begin{axis}[
    xmode=log,
    log basis x=10,
    xtick={0.25,0.5,1,2,4,8,16,32},
    xticklabels={0.25,0.5,1,2,4,8,16,32},
    xmin=0.25, xmax=32,
    y dir=reverse,
    ymin=1, ymax=4,
    width=0.8\linewidth,
    height=3.5cm,
    axis lines*=left,
    xlabel style = {align=center, font=\footnotesize},
    xlabel={\textcolor{ppt-blue}{$\xleftarrow{\textrm{More Likely Crew}}$}\hspace{1.5cm}\textcolor{ppt-red}{$\xrightarrow{\textrm{More Likely Imposter}}$} \\ Odds Ratio },
    ylabel={Speech Act},
    ytick={1,2,3,4},
    yticklabels={
        Representatives,
        Directives,
        Commissives,
        Expressives
    },
    tick align=outside,
    enlarge y limits=0.2,
    xticklabel style={
      /pgf/number format/fixed,
      /pgf/number format/precision=2
    },
    scaled x ticks=false,
]

\addplot[black, dashed] coordinates {(1,0) (1,5)};

\addplot+[only marks, mark=o, mark size=2pt] 
coordinates {
    (3.36, 1)  
    (0.37, 2)  
    (7.18, 3)  
    (0.91, 4)  
};

\addplot+[thick, black] coordinates {(2.60,1) (4.34,1)};   
\addplot+[thick, black] coordinates {(0.29,2) (0.46,2)};   
\addplot+[thick, black] coordinates {(1.39,3) (37.04,3)};  
\addplot+[thick, black] coordinates {(0.50,4) (1.67,4)};   

\end{axis}
\end{tikzpicture}

%% file: figs/rq2_logit.tex
\begin{tikzpicture}
\begin{axis}[
    xmode=log,
    log basis x=10,
    xtick={0.5,0.75,1,1.25,1.5,2},
    xticklabels={0.5,0.75,1,1.25,1.5,2},
    xmin=0.5, xmax=2.0,
    y dir=reverse,
    ymin=1, ymax=3,
    width=0.8\linewidth,
    height=3.15cm,
    axis lines*=left,
    xlabel style = {align=center, font=\footnotesize},
    xlabel={\textcolor{ppt-red}{$\xleftarrow{\textrm{Imposter Win}}$}\hspace{1.5cm}\textcolor{ppt-blue}{$\xrightarrow{\textrm{Crew Win}}$} \\ Odds Ratio },
    ylabel={Speech Act},
    ytick={1,2,3},
    yticklabels={Directives, Representatives, Commissives},
    tick align=outside,
    enlarge y limits=0.2,
]

\addplot[black, dashed] coordinates {(1,0) (1,4)};

\addplot+[only marks, mark=o, mark size=2pt] 
coordinates {
    (1.34,1)   
    (1.04,2)   
    (0.79,3)   
};

\addplot+[thick, black] coordinates {(1.17,1) (1.55,1)};   
\addplot+[thick, black] coordinates {(0.93,2) (1.16,2)};   
\addplot+[thick, black] coordinates {(0.64,3) (0.98,3)};   

\end{axis}
\end{tikzpicture}

%% file: figs/rq3_deception.tex
\pgfplotscreateplotcyclelist{imposter}{
    ppt-red, mark=diamond\\
}
\begin{tikzpicture}
\begin{groupplot}[
  group style={group size=1 by 2, vertical sep=2.0cm},
  width=0.80\linewidth,
  height=4cm,
  xlabel={Deception type},
  ylabel={Logit proportion},
  xtick={1,2,3,4},
  ymin=-8, ymax=4,
  xticklabels={Concealment, Falsification, Equivocation, Missing},
  ytick={-13.8,-9.2,-6.9,-4.6,-2.2,0,2.2,4.6},
  yticklabels={0.000001,0.0001,0.001,0.01,0.1,0.5,0.9,0.99},
  ymajorgrids=true,
  legend style={at={(1.02,0.5)}, anchor=west, legend columns=1}
]

\nextgroupplot[title={(a) Winning vs. Losing Games}, title style={font=\small, yshift=-3mm}, cycle list name=winloss,]

\addplot+[only marks, mark=o, fill=gray!40,
          error bars/.cd, y dir=both, y explicit]
coordinates {
  (1.00, -5.016) +- (0, 0.175)
  (2.00, -3.760) +- (0, 0.095)
  (3.00, 2.292) +- (0, 0.049)
  (4.00, -2.709) +- (0, 0.058)
};
\addlegendentry{Losing}

\addplot+[only marks, mark=o, fill=gray!10,
          error bars/.cd, y dir=both, y explicit]
coordinates {
  (1.15, -5.298) +- (0, 0.173)
  (2.15, -4.074) +- (0, 0.092)
  (3.15,  2.638) +- (0, 0.048)
  (4.15, -3.118) +- (0, 0.059)
};
\addlegendentry{Winning}

\nextgroupplot[title={(b) Imposter }, title style={font=\small, yshift=-3mm}, cycle list name=imposter,]

\addplot+[only marks, mark=o, fill=gray!10,
          error bars/.cd, y dir=both, y explicit]
coordinates {
  (1.0, -5.029) +- (0, 0.175)
  (2.0, -3.807) +- (0, 0.097)
  (3.0, 2.338) +- (0, 0.050)
  (4.0, -2.755) +- (0, 0.059)
};
\addlegendentry{Impostor}

\end{groupplot}
\end{tikzpicture}

%% file: main.bbl
\begin{thebibliography}{40}


\ifx \showCODEN    \undefined \def \showCODEN     #1{\unskip}     \fi
\ifx \showDOI      \undefined \def \showDOI       #1{#1}\fi
\ifx \showISBNx    \undefined \def \showISBNx     #1{\unskip}     \fi
\ifx \showISBNxiii \undefined \def \showISBNxiii  #1{\unskip}     \fi
\ifx \showISSN     \undefined \def \showISSN      #1{\unskip}     \fi
\ifx \showLCCN     \undefined \def \showLCCN      #1{\unskip}     \fi
\ifx \shownote     \undefined \def \shownote      #1{#1}          \fi
\ifx \showarticletitle \undefined \def \showarticletitle #1{#1}   \fi
\ifx \showURL      \undefined \def \showURL       {\relax}        \fi
\providecommand\bibfield[2]{#2}
\providecommand\bibinfo[2]{#2}
\providecommand\natexlab[1]{#1}
\providecommand\showeprint[2][]{arXiv:#2}

\bibitem[\protect\citeauthoryear{Adhikari, Gmytrasiewicz, Wang, Falcone, and Zhang}{Adhikari et~al\mbox{.}}{2021}]%
        {adhikari2021telling}
\bibfield{author}{\bibinfo{person}{Sarit Adhikari}, \bibinfo{person}{Piotr~J Gmytrasiewicz}, \bibinfo{person}{D Wang}, \bibinfo{person}{R Falcone}, {and} \bibinfo{person}{J Zhang}.} \bibinfo{year}{2021}\natexlab{}.
\newblock \showarticletitle{Telling Friend from Foe-Towards a Bayesian Approach to Sincerity and Deception.}. In \bibinfo{booktitle}{\emph{TRUST@ AAMAS}}.
\newblock


\bibitem[\protect\citeauthoryear{Austin}{Austin}{1975}]%
        {austin1975things}
\bibfield{author}{\bibinfo{person}{John~Langshaw Austin}.} \bibinfo{year}{1975}\natexlab{}.
\newblock \bibinfo{booktitle}{\emph{How to do things with words}}.
\newblock \bibinfo{publisher}{Harvard university press}.
\newblock


\bibitem[\protect\citeauthoryear{Bai, Jones, Ndousse, Askell, Chen, DasSarma, Drain, Fort, Ganguli, Henighan, et~al\mbox{.}}{Bai et~al\mbox{.}}{2022}]%
        {bai2022training}
\bibfield{author}{\bibinfo{person}{Yuntao Bai}, \bibinfo{person}{Andy Jones}, \bibinfo{person}{Kamal Ndousse}, \bibinfo{person}{Amanda Askell}, \bibinfo{person}{Anna Chen}, \bibinfo{person}{Nova DasSarma}, \bibinfo{person}{Dawn Drain}, \bibinfo{person}{Stanislav Fort}, \bibinfo{person}{Deep Ganguli}, \bibinfo{person}{Tom Henighan}, {et~al\mbox{.}}} \bibinfo{year}{2022}\natexlab{}.
\newblock \showarticletitle{Training a helpful and harmless assistant with reinforcement learning from human feedback}.
\newblock \bibinfo{journal}{\emph{arXiv preprint arXiv:2204.05862}} (\bibinfo{year}{2022}).
\newblock


\bibitem[\protect\citeauthoryear{Bellifemine, Poggi, and Rimassa}{Bellifemine et~al\mbox{.}}{2000}]%
        {bellifemine2000developing}
\bibfield{author}{\bibinfo{person}{Fabio Bellifemine}, \bibinfo{person}{Agostino Poggi}, {and} \bibinfo{person}{Giovanni Rimassa}.} \bibinfo{year}{2000}\natexlab{}.
\newblock \showarticletitle{Developing multi-agent systems with JADE}. In \bibinfo{booktitle}{\emph{International workshop on agent theories, architectures, and languages}}. Springer, \bibinfo{pages}{89--103}.
\newblock


\bibitem[\protect\citeauthoryear{Buller and Burgoon}{Buller and Burgoon}{1996}]%
        {buller1996interpersonal}
\bibfield{author}{\bibinfo{person}{David~B Buller} {and} \bibinfo{person}{Judee~K Burgoon}.} \bibinfo{year}{1996}\natexlab{}.
\newblock \showarticletitle{Interpersonal deception theory}.
\newblock \bibinfo{journal}{\emph{Communication theory}} \bibinfo{volume}{6}, \bibinfo{number}{3} (\bibinfo{year}{1996}), \bibinfo{pages}{203--242}.
\newblock


\bibitem[\protect\citeauthoryear{Chern, Hu, Yang, Chern, Guo, Jin, Wang, and Liu}{Chern et~al\mbox{.}}{2024}]%
        {chern2024behonest}
\bibfield{author}{\bibinfo{person}{Steffi Chern}, \bibinfo{person}{Zhulin Hu}, \bibinfo{person}{Yuqing Yang}, \bibinfo{person}{Ethan Chern}, \bibinfo{person}{Yuan Guo}, \bibinfo{person}{Jiahe Jin}, \bibinfo{person}{Binjie Wang}, {and} \bibinfo{person}{Pengfei Liu}.} \bibinfo{year}{2024}\natexlab{}.
\newblock \showarticletitle{Behonest: Benchmarking honesty in large language models}.
\newblock \bibinfo{journal}{\emph{arXiv preprint arXiv:2406.13261}} (\bibinfo{year}{2024}).
\newblock


\bibitem[\protect\citeauthoryear{Christian and Young}{Christian and Young}{2004}]%
        {christian2004strategic}
\bibfield{author}{\bibinfo{person}{David Christian} {and} \bibinfo{person}{R~Michael Young}.} \bibinfo{year}{2004}\natexlab{}.
\newblock \showarticletitle{Strategic deception in agents}. In \bibinfo{booktitle}{\emph{Autonomous Agents and Multiagent Systems, International Joint Conference on}}, Vol.~\bibinfo{volume}{2}. IEEE Computer Society, \bibinfo{pages}{218--226}.
\newblock


\bibitem[\protect\citeauthoryear{(FAIR)†, Bakhtin, Brown, Dinan, Farina, Flaherty, Fried, Goff, Gray, Hu, et~al\mbox{.}}{(FAIR)† et~al\mbox{.}}{2022}]%
        {meta2022human}
\bibfield{author}{\bibinfo{person}{Meta Fundamental AI Research Diplomacy~Team (FAIR)†}, \bibinfo{person}{Anton Bakhtin}, \bibinfo{person}{Noam Brown}, \bibinfo{person}{Emily Dinan}, \bibinfo{person}{Gabriele Farina}, \bibinfo{person}{Colin Flaherty}, \bibinfo{person}{Daniel Fried}, \bibinfo{person}{Andrew Goff}, \bibinfo{person}{Jonathan Gray}, \bibinfo{person}{Hengyuan Hu}, {et~al\mbox{.}}} \bibinfo{year}{2022}\natexlab{}.
\newblock \showarticletitle{Human-level play in the game of Diplomacy by combining language models with strategic reasoning}.
\newblock \bibinfo{journal}{\emph{Science}} \bibinfo{volume}{378}, \bibinfo{number}{6624} (\bibinfo{year}{2022}), \bibinfo{pages}{1067--1074}.
\newblock


\bibitem[\protect\citeauthoryear{Finin, Fritzson, McKay, and McEntire}{Finin et~al\mbox{.}}{1994}]%
        {finin1994kqml}
\bibfield{author}{\bibinfo{person}{Tim Finin}, \bibinfo{person}{Richard Fritzson}, \bibinfo{person}{Don McKay}, {and} \bibinfo{person}{Robin McEntire}.} \bibinfo{year}{1994}\natexlab{}.
\newblock \showarticletitle{KQML as an agent communication language}. In \bibinfo{booktitle}{\emph{Proceedings of the third international conference on Information and knowledge management}}. \bibinfo{pages}{456--463}.
\newblock


\bibitem[\protect\citeauthoryear{FIPA}{FIPA}{2009}]%
        {fipa2009fipa}
\bibfield{author}{\bibinfo{person}{ACL FIPA}.} \bibinfo{year}{2009}\natexlab{}.
\newblock \bibinfo{title}{FIPA Agent Communication Language Specifications}.
\newblock \bibinfo{howpublished}{Available at http://www.fipa.org/specs/}.
\newblock


\bibitem[\protect\citeauthoryear{Goldowsky-Dill, Chughtai, Heimersheim, and Hobbhahn}{Goldowsky-Dill et~al\mbox{.}}{2025}]%
        {goldowsky2025detecting}
\bibfield{author}{\bibinfo{person}{Nicholas Goldowsky-Dill}, \bibinfo{person}{Bilal Chughtai}, \bibinfo{person}{Stefan Heimersheim}, {and} \bibinfo{person}{Marius Hobbhahn}.} \bibinfo{year}{2025}\natexlab{}.
\newblock \showarticletitle{Detecting Strategic Deception Using Linear Probes}.
\newblock \bibinfo{journal}{\emph{arXiv preprint arXiv:2502.03407}} (\bibinfo{year}{2025}).
\newblock


\bibitem[\protect\citeauthoryear{Golechha and Garriga-Alonso}{Golechha and Garriga-Alonso}{2025}]%
        {golechha2025amongus}
\bibfield{author}{\bibinfo{person}{Satvik Golechha} {and} \bibinfo{person}{Adri{\`a} Garriga-Alonso}.} \bibinfo{year}{2025}\natexlab{}.
\newblock \showarticletitle{Among us: A sandbox for measuring and detecting agentic deception}.
\newblock \bibinfo{journal}{\emph{arXiv preprint arXiv:2504.04072}} (\bibinfo{year}{2025}).
\newblock


\bibitem[\protect\citeauthoryear{Greenblatt, Denison, Wright, Roger, MacDiarmid, Marks, Treutlein, Belonax, Chen, Duvenaud, et~al\mbox{.}}{Greenblatt et~al\mbox{.}}{2024}]%
        {greenblatt2024alignment}
\bibfield{author}{\bibinfo{person}{Ryan Greenblatt}, \bibinfo{person}{Carson Denison}, \bibinfo{person}{Benjamin Wright}, \bibinfo{person}{Fabien Roger}, \bibinfo{person}{Monte MacDiarmid}, \bibinfo{person}{Sam Marks}, \bibinfo{person}{Johannes Treutlein}, \bibinfo{person}{Tim Belonax}, \bibinfo{person}{Jack Chen}, \bibinfo{person}{David Duvenaud}, {et~al\mbox{.}}} \bibinfo{year}{2024}\natexlab{}.
\newblock \showarticletitle{Alignment faking in large language models}.
\newblock \bibinfo{journal}{\emph{arXiv preprint arXiv:2412.14093}} (\bibinfo{year}{2024}).
\newblock


\bibitem[\protect\citeauthoryear{Hagendorff}{Hagendorff}{2024}]%
        {hagendorff2024deception}
\bibfield{author}{\bibinfo{person}{Thilo Hagendorff}.} \bibinfo{year}{2024}\natexlab{}.
\newblock \showarticletitle{Deception abilities emerged in large language models}.
\newblock \bibinfo{journal}{\emph{Proceedings of the National Academy of Sciences}} \bibinfo{volume}{121}, \bibinfo{number}{24} (\bibinfo{year}{2024}), \bibinfo{pages}{e2317967121}.
\newblock


\bibitem[\protect\citeauthoryear{Hubinger, Denison, Mu, Lambert, Tong, MacDiarmid, Lanham, Ziegler, Maxwell, Cheng, et~al\mbox{.}}{Hubinger et~al\mbox{.}}{2024}]%
        {hubinger2024sleeper}
\bibfield{author}{\bibinfo{person}{E Hubinger}, \bibinfo{person}{C Denison}, \bibinfo{person}{J Mu}, \bibinfo{person}{M Lambert}, \bibinfo{person}{M Tong}, \bibinfo{person}{M MacDiarmid}, \bibinfo{person}{T Lanham}, \bibinfo{person}{DM Ziegler}, \bibinfo{person}{T Maxwell}, \bibinfo{person}{N Cheng}, {et~al\mbox{.}}} \bibinfo{year}{2024}\natexlab{}.
\newblock \showarticletitle{Sleeper agents: Training deceptive llms that persist through safety training, 2024}.
\newblock \bibinfo{journal}{\emph{arXiv preprint arXiv:2401.05566}} (\bibinfo{year}{2024}).
\newblock


\bibitem[\protect\citeauthoryear{Huynh, Cao, Wu, et~al\mbox{.}}{Huynh et~al\mbox{.}}{2024}]%
        {huynh2024multi}
\bibfield{author}{\bibinfo{person}{Nhat-Minh Huynh}, \bibinfo{person}{Hoang-Giang Cao}, \bibinfo{person}{I Wu}, {et~al\mbox{.}}} \bibinfo{year}{2024}\natexlab{}.
\newblock \showarticletitle{Multi-Agent Training for Pommerman: Curriculum Learning and Population-based Self-Play Approach}.
\newblock \bibinfo{journal}{\emph{arXiv preprint arXiv:2407.00662}} (\bibinfo{year}{2024}).
\newblock


\bibitem[\protect\citeauthoryear{Jahan and Mell}{Jahan and Mell}{2025}]%
        {jahan2025decoding}
\bibfield{author}{\bibinfo{person}{Nusrath Jahan} {and} \bibinfo{person}{Johnathan Mell}.} \bibinfo{year}{2025}\natexlab{}.
\newblock \showarticletitle{Decoding Negotiation Dynamics: The Impact of Opponent Identity and Privacy on Strategy, Deception, and Emotional Transparency in Human-Agent Interaction}. In \bibinfo{booktitle}{\emph{Proceedings of the 24th International Conference on Autonomous Agents and Multiagent Systems}}. \bibinfo{pages}{2562--2564}.
\newblock


\bibitem[\protect\citeauthoryear{Jones and Bergen}{Jones and Bergen}{2024}]%
        {jones2024lies}
\bibfield{author}{\bibinfo{person}{Cameron~R Jones} {and} \bibinfo{person}{Benjamin~K Bergen}.} \bibinfo{year}{2024}\natexlab{}.
\newblock \showarticletitle{Lies, damned lies, and distributional language statistics: Persuasion and deception with large language models}.
\newblock \bibinfo{journal}{\emph{arXiv preprint arXiv:2412.17128}} (\bibinfo{year}{2024}).
\newblock


\bibitem[\protect\citeauthoryear{Karten, Kailas, Li, and Sycara}{Karten et~al\mbox{.}}{2023}]%
        {karten2023role}
\bibfield{author}{\bibinfo{person}{Seth Karten}, \bibinfo{person}{Siva Kailas}, \bibinfo{person}{Huao Li}, {and} \bibinfo{person}{Katia Sycara}.} \bibinfo{year}{2023}\natexlab{}.
\newblock \showarticletitle{On the role of emergent communication for social learning in multi-agent reinforcement learning}.
\newblock \bibinfo{journal}{\emph{arXiv preprint arXiv:2302.14276}} (\bibinfo{year}{2023}).
\newblock


\bibitem[\protect\citeauthoryear{Kraus}{Kraus}{1997}]%
        {kraus1997negotiation}
\bibfield{author}{\bibinfo{person}{Sarit Kraus}.} \bibinfo{year}{1997}\natexlab{}.
\newblock \showarticletitle{Negotiation and cooperation in multi-agent environments}.
\newblock \bibinfo{journal}{\emph{Artificial intelligence}} \bibinfo{volume}{94}, \bibinfo{number}{1-2} (\bibinfo{year}{1997}), \bibinfo{pages}{79--97}.
\newblock


\bibitem[\protect\citeauthoryear{Labrou and Finin}{Labrou and Finin}{1997}]%
        {labrou1997semantics}
\bibfield{author}{\bibinfo{person}{Yannis Labrou} {and} \bibinfo{person}{Tim Finin}.} \bibinfo{year}{1997}\natexlab{}.
\newblock \showarticletitle{Semantics for an agent communication language}. In \bibinfo{booktitle}{\emph{International Workshop on Agent Theories, Architectures, and Languages}}. Springer, \bibinfo{pages}{209--214}.
\newblock


\bibitem[\protect\citeauthoryear{Li, Nourkhiz~Mahjoub, Chalaki, Tadiparthi, Lee, Moradi~Pari, Lewis, and Sycara}{Li et~al\mbox{.}}{2024}]%
        {li2024language}
\bibfield{author}{\bibinfo{person}{Huao Li}, \bibinfo{person}{Hossein Nourkhiz~Mahjoub}, \bibinfo{person}{Behdad Chalaki}, \bibinfo{person}{Vaishnav Tadiparthi}, \bibinfo{person}{Kwonjoon Lee}, \bibinfo{person}{Ehsan Moradi~Pari}, \bibinfo{person}{Charles Lewis}, {and} \bibinfo{person}{Katia Sycara}.} \bibinfo{year}{2024}\natexlab{}.
\newblock \showarticletitle{Language grounded multi-agent reinforcement learning with human-interpretable communication}.
\newblock \bibinfo{journal}{\emph{Advances in Neural Information Processing Systems}}  \bibinfo{volume}{37} (\bibinfo{year}{2024}), \bibinfo{pages}{87908--87933}.
\newblock


\bibitem[\protect\citeauthoryear{Mayoral-Macau}{Mayoral-Macau}{2025}]%
        {mayoral2025environment}
\bibfield{author}{\bibinfo{person}{Arnau Mayoral-Macau}.} \bibinfo{year}{2025}\natexlab{}.
\newblock \showarticletitle{Environment-Centered Design of Ethical Environments}. In \bibinfo{booktitle}{\emph{Proceedings of the 24th International Conference on Autonomous Agents and Multiagent Systems}}. \bibinfo{pages}{2953--2955}.
\newblock


\bibitem[\protect\citeauthoryear{Nguyen, Le, Do, Gupta, Venkatesh, and Tran}{Nguyen et~al\mbox{.}}{2025}]%
        {nguyen2025navigating}
\bibfield{author}{\bibinfo{person}{Dung Nguyen}, \bibinfo{person}{Hung Le}, \bibinfo{person}{Kien Do}, \bibinfo{person}{Sunil Gupta}, \bibinfo{person}{Svetha Venkatesh}, {and} \bibinfo{person}{Truyen Tran}.} \bibinfo{year}{2025}\natexlab{}.
\newblock \showarticletitle{Navigating Social Dilemmas with LLM-based Agents via Consideration of Future Consequences}. In \bibinfo{booktitle}{\emph{Proceedings of the 24th International Conference on Autonomous Agents and Multiagent Systems}}. \bibinfo{pages}{2693--2695}.
\newblock


\bibitem[\protect\citeauthoryear{O'Gara}{O'Gara}{2023}]%
        {o2023hoodwinked}
\bibfield{author}{\bibinfo{person}{Aidan O'Gara}.} \bibinfo{year}{2023}\natexlab{}.
\newblock \showarticletitle{Hoodwinked: Deception and cooperation in a text-based game for language models}.
\newblock \bibinfo{journal}{\emph{arXiv preprint arXiv:2308.01404}} (\bibinfo{year}{2023}).
\newblock


\bibitem[\protect\citeauthoryear{Okasha}{Okasha}{2006}]%
        {okasha2006evolution}
\bibfield{author}{\bibinfo{person}{Samir Okasha}.} \bibinfo{year}{2006}\natexlab{}.
\newblock \bibinfo{booktitle}{\emph{Evolution and the levels of selection}}.
\newblock \bibinfo{publisher}{Clarendon Press}.
\newblock


\bibitem[\protect\citeauthoryear{Ouyang, Wu, Jiang, Almeida, Wainwright, Mishkin, Zhang, Agarwal, Slama, Ray, et~al\mbox{.}}{Ouyang et~al\mbox{.}}{2022}]%
        {ouyang2022training}
\bibfield{author}{\bibinfo{person}{Long Ouyang}, \bibinfo{person}{Jeffrey Wu}, \bibinfo{person}{Xu Jiang}, \bibinfo{person}{Diogo Almeida}, \bibinfo{person}{Carroll Wainwright}, \bibinfo{person}{Pamela Mishkin}, \bibinfo{person}{Chong Zhang}, \bibinfo{person}{Sandhini Agarwal}, \bibinfo{person}{Katarina Slama}, \bibinfo{person}{Alex Ray}, {et~al\mbox{.}}} \bibinfo{year}{2022}\natexlab{}.
\newblock \showarticletitle{Training language models to follow instructions with human feedback}.
\newblock \bibinfo{journal}{\emph{Advances in neural information processing systems}}  \bibinfo{volume}{35} (\bibinfo{year}{2022}), \bibinfo{pages}{27730--27744}.
\newblock


\bibitem[\protect\citeauthoryear{Panait and Luke}{Panait and Luke}{2005}]%
        {panait2005cooperative}
\bibfield{author}{\bibinfo{person}{Liviu Panait} {and} \bibinfo{person}{Sean Luke}.} \bibinfo{year}{2005}\natexlab{}.
\newblock \showarticletitle{Cooperative multi-agent learning: The state of the art}.
\newblock \bibinfo{journal}{\emph{Autonomous agents and multi-agent systems}} \bibinfo{volume}{11}, \bibinfo{number}{3} (\bibinfo{year}{2005}), \bibinfo{pages}{387--434}.
\newblock


\bibitem[\protect\citeauthoryear{Rahwan}{Rahwan}{2005}]%
        {rahwan2005guest}
\bibfield{author}{\bibinfo{person}{Iyad Rahwan}.} \bibinfo{year}{2005}\natexlab{}.
\newblock \showarticletitle{Guest editorial: Argumentation in multi-agent systems}.
\newblock \bibinfo{journal}{\emph{Autonomous Agents and Multi-Agent Systems}} \bibinfo{volume}{11}, \bibinfo{number}{2} (\bibinfo{year}{2005}), \bibinfo{pages}{115--125}.
\newblock


\bibitem[\protect\citeauthoryear{Sarkadi and Lewis}{Sarkadi and Lewis}{2024}]%
        {sarkadi2024triangles}
\bibfield{author}{\bibinfo{person}{{\c{S}}tefan Sarkadi} {and} \bibinfo{person}{Peter~R Lewis}.} \bibinfo{year}{2024}\natexlab{}.
\newblock \showarticletitle{The triangles of dishonesty: modelling the evolution of lies, bullshit, and deception in agent societies}. In \bibinfo{booktitle}{\emph{Proc. of the 23rd International Conference on Autonomous Agents and Multiagent Systems (AAMAS 2024)}}. International Foundation for Autonomous Agents and Multiagent Systems (IFAAMAS).
\newblock


\bibitem[\protect\citeauthoryear{Sarkadi, Mei, and Awad}{Sarkadi et~al\mbox{.}}{2023}]%
        {sarkadi2023should}
\bibfield{author}{\bibinfo{person}{Stefan Sarkadi}, \bibinfo{person}{Peidong Mei}, {and} \bibinfo{person}{Edmond Awad}.} \bibinfo{year}{2023}\natexlab{}.
\newblock \showarticletitle{Should My Agent Lie for Me? Public Moral Perspectives on Deceptive AI}. In \bibinfo{booktitle}{\emph{International Conference on Autonomous Agents and Multiagent Systems}}. Springer, \bibinfo{pages}{151--179}.
\newblock


\bibitem[\protect\citeauthoryear{Sarkar, Xia, Liu, and Sadigh}{Sarkar et~al\mbox{.}}{2025}]%
        {sarkar2025training}
\bibfield{author}{\bibinfo{person}{Bidipta Sarkar}, \bibinfo{person}{Warren Xia}, \bibinfo{person}{C~Karen Liu}, {and} \bibinfo{person}{Dorsa Sadigh}.} \bibinfo{year}{2025}\natexlab{}.
\newblock \showarticletitle{Training language models for social deduction with multi-agent reinforcement learning}.
\newblock \bibinfo{journal}{\emph{arXiv preprint arXiv:2502.06060}} (\bibinfo{year}{2025}).
\newblock


\bibitem[\protect\citeauthoryear{Searle}{Searle}{1969}]%
        {searle1969speech}
\bibfield{author}{\bibinfo{person}{John~R Searle}.} \bibinfo{year}{1969}\natexlab{}.
\newblock \showarticletitle{Speech acts: An essay in the philosophy of language}.
\newblock \bibinfo{journal}{\emph{Cambridge University}} (\bibinfo{year}{1969}).
\newblock


\bibitem[\protect\citeauthoryear{Shen}{Shen}{2025}]%
        {shen2025emergent}
\bibfield{author}{\bibinfo{person}{Wenjie Shen}.} \bibinfo{year}{2025}\natexlab{}.
\newblock \showarticletitle{Emergent Language in Multi-Agent Systems: A Multi-Task Learning Approach}. In \bibinfo{booktitle}{\emph{Proceedings of the 2025 International Conference on Big Data and Informatization Education}}. \bibinfo{pages}{539--544}.
\newblock


\bibitem[\protect\citeauthoryear{Stone and Veloso}{Stone and Veloso}{2000}]%
        {stone2000multiagent}
\bibfield{author}{\bibinfo{person}{Peter Stone} {and} \bibinfo{person}{Manuela Veloso}.} \bibinfo{year}{2000}\natexlab{}.
\newblock \showarticletitle{Multiagent systems: A survey from a machine learning perspective}.
\newblock \bibinfo{journal}{\emph{Autonomous Robots}} \bibinfo{volume}{8}, \bibinfo{number}{3} (\bibinfo{year}{2000}), \bibinfo{pages}{345--383}.
\newblock


\bibitem[\protect\citeauthoryear{Vrij}{Vrij}{2008}]%
        {vrij2008detecting}
\bibfield{author}{\bibinfo{person}{Aldert Vrij}.} \bibinfo{year}{2008}\natexlab{}.
\newblock \bibinfo{booktitle}{\emph{Detecting Lies and Deceit: Pitfalls and Opportunities}}.
\newblock \bibinfo{publisher}{Wiley}.
\newblock


\bibitem[\protect\citeauthoryear{Wang, Liu, Zheng, Qi, Chen, Yang, Zhao, Wang, Song, and Huang}{Wang et~al\mbox{.}}{2023}]%
        {wang2023avalon}
\bibfield{author}{\bibinfo{person}{Shenzhi Wang}, \bibinfo{person}{Chang Liu}, \bibinfo{person}{Zilong Zheng}, \bibinfo{person}{Siyuan Qi}, \bibinfo{person}{Shuo Chen}, \bibinfo{person}{Qisen Yang}, \bibinfo{person}{Andrew Zhao}, \bibinfo{person}{Chaofei Wang}, \bibinfo{person}{Shiji Song}, {and} \bibinfo{person}{Gao Huang}.} \bibinfo{year}{2023}\natexlab{}.
\newblock \showarticletitle{Avalon's game of thoughts: Battle against deception through recursive contemplation}.
\newblock \bibinfo{journal}{\emph{arXiv preprint arXiv:2310.01320}} (\bibinfo{year}{2023}).
\newblock


\bibitem[\protect\citeauthoryear{Ward, Toni, and Belardinelli}{Ward et~al\mbox{.}}{2023}]%
        {ward2023defining}
\bibfield{author}{\bibinfo{person}{Francis~Rhys Ward}, \bibinfo{person}{Francesca Toni}, {and} \bibinfo{person}{Francesco Belardinelli}.} \bibinfo{year}{2023}\natexlab{}.
\newblock \showarticletitle{Defining deception in structural causal games}. In \bibinfo{booktitle}{\emph{Proceedings of the 2023 International Conference on Autonomous Agents and Multiagent Systems}}. \bibinfo{pages}{2902--2904}.
\newblock


\bibitem[\protect\citeauthoryear{Wooldridge}{Wooldridge}{2009}]%
        {wooldridge2009introduction}
\bibfield{author}{\bibinfo{person}{Michael Wooldridge}.} \bibinfo{year}{2009}\natexlab{}.
\newblock \bibinfo{booktitle}{\emph{An introduction to multiagent systems}}.
\newblock \bibinfo{publisher}{John wiley \& sons}.
\newblock


\bibitem[\protect\citeauthoryear{Xu, Wang, Li, Luo, Wang, Liu, and Liu}{Xu et~al\mbox{.}}{2023}]%
        {xu2023exploring}
\bibfield{author}{\bibinfo{person}{Yuzhuang Xu}, \bibinfo{person}{Shuo Wang}, \bibinfo{person}{Peng Li}, \bibinfo{person}{Fuwen Luo}, \bibinfo{person}{Xiaolong Wang}, \bibinfo{person}{Weidong Liu}, {and} \bibinfo{person}{Yang Liu}.} \bibinfo{year}{2023}\natexlab{}.
\newblock \showarticletitle{Exploring large language models for communication games: An empirical study on werewolf}.
\newblock \bibinfo{journal}{\emph{arXiv preprint arXiv:2309.04658}} (\bibinfo{year}{2023}).
\newblock


\end{thebibliography}
